\documentstyle[preprint,epsf,%
aps]{revtex}
\tighten
\firstfigfalse
\begin{document}
\preprint{{\tt hep-ph/9408262 (rev.)} / DESY 94-132 }
\title{NONABELIAN DEBYE SCREENING
    \\ IN ONE-LOOP RESUMMED PERTURBATION THEORY }
\author{Anton K. Rebhan\thanks{On leave of absence from
	Institut f\"ur Theoretische Physik der
         Technischen Universit\"at Wien}}
\address{DESY, Gruppe Theorie,\\
	Notkestra\ss e 85, D-22603 Hamburg, Germany}

\author{In memory of Tanguy Altherr}

\maketitle

\begin{abstract}
Debye screening of static chromoelectric fields at high temperature
is investigated at next-to-leading order through one-loop resummed
perturbation theory. At this order the gluon propagator appears to
give rise to strong deviations from a Yukawa form of screening.
Generally,
an oscillatory behavior is found which asymptotically becomes
repulsive,
but in a gauge-dependent manner. However, these features are strongly
sensitive to the existence of screening of static magnetic fields. It
is shown that a small magnetic screening mass can restore exponential
screening with a gauge independent value of the screening mass,
which depends
logarithmically on the magnitude of the magnetic mass. Recent results
obtained in temporal axial gauge, which instead indicate an asymptotic
(repulsive) power-law behaviour of screening, are also critically
discussed.
In order to arrive at a gauge-invariant treatment of chromoelectric
screening, Polyakov loop correlations are considered, both with and
without dynamical gauge symmetry breaking. Again a crucial sensitivity
to the scale of magnetic screening is found. A detailed comparison
of the
perturbative results with recent high-precision lattice simulations
of the
SU(2) Polyakov loop correlator is made, which are found to
agree well with the perturbative result in the symmetric phase
when a magnetic mass $\sim g^2T/4$ is included.
\end{abstract}


\section{Introduction and summary}
It is well established that
quantum chromodynamics (QCD) at sufficiently high temperature and/or
density is in a deconfined phase. Although the coupling $g$ 
remains uncomfortably large up to astronomically
large energy scales, it is
hoped that this regime may be accessible by perturbative quantum
field theory at finite temperature and density. Indeed, perturbation
theory works reasonably well at the energy scale set by the
temperature,
but it runs into infrared singularities when probing the softer
scale $gT$, where a quasiparticle picture becomes relevant.

The leading-order dispersion laws of the quasi-particle
excitations are
determined by the high-temperature limit of one-loop Feynman diagrams,
the so-called ``hard thermal loops'' (HTL)\cite{FT,BP},
which can be understood
in classical terms\cite{Silin,forwscatt}. But already at
next-to-leading order the dispersion laws receive contributions from
all orders of the conventional perturbation series. It
has been shown in particular by Braaten and Pisarski\cite{BP}
that an accurate and gauge-independent
calculation of corrections to the HTL dispersion laws\cite{Damping,HS}
requires
an improved perturbation theory which resums HTL contributions.
This need is generic in perturbative thermal field
theory\cite{Parwani,SED}, but in nonabelian gauge theories
there is another, pernicious barrier for perturbation theory.
Static magnetic
fields are not screened at the HTL level, and the self-interactions
of these lead to a breakdown of perturbation theory at a certain loop
order depending on the quantity under consideration.
The corresponding infra-red singularities are commonly expected to be
cured by the dynamical generation of a magnetic screening mass
$\sim g^2T$, but its nature is still unclear.
By superficial infra-red power counting, the
next-to-leading order corrections to the dispersion laws
are still below this critical order, however mass-shell singularities
tend to generate a logarithmic sensitivity to the magnetic mass
scale\cite{Damping2,Damping3}.

In particular, such a sensitivity to the magnetic scale is found in
a next-to-leading order calculation of the chromoelectric screening
mass\cite{AKR}, which will be recapitulated in Sect.~2. Such a
nonabelian analogue of the classical Debye mass is supposed to
account for an exponential screening of colour charges, and this will
presumably provide an important characteristics of
the hypothetical quark-gluon plasma\cite{Satz}.
There may however be substantial corrections to the pre-exponential
part of the screening function, or even non-exponential behaviour
on sufficiently large distances. For instance, in an electron gas
it has been found\cite{CM}
that quantum corrections give rise to an asymptotic
$1/r^{10}$ behaviour at order $\hbar^4$, which eventually supersedes
exponential screening.

The main objective of this paper
will be the discussion of the various possibilities to define
screening functions and of the results
found at next-to-leading order. The simplest possibility is to
inspect the chromoeletric field induced by a weak static colour source,
which is however a gauge dependent quantity.
In Sect.~3
the covariant gauge result is analysed and, if taken at
face value, it indicates a strong departure from the
expected Yukawa-type potential. The pre-exponential screening
function oscillates 
and approaches
a negative value asymptotically,
signalling a repulsive exponential tail.
However, the analytic structures which gives rise to these phenomena
turn out to be strongly sensitive to the existence of a
magnetic screening mass, and are in fact found to be largely tamed by
the latter. A small magnetic mass can restore exponential screening
with a gauge independent Debye mass.
In Sect.~4 recent results\cite{BK,PW} obtained for the temporal
axial gauge are discussed, which qualitatively differ from the
covariant gauge results and which seem to imply an asymptotic
{\em repulsive power-law} behaviour in place of exponential screening.
The motivation for using the notoriously troublesome temporal gauge
at finite temperature is its relation to the correlation of
two chromoelectric field strength operators. Evaluating the latter
in covariant gauge at next-to-leading order reveals a gauge dependence
which makes it clear that one cannot attribute direct physical
meaning to it either. Moreover it is argued that the different analytic
structure which is responsible for the different asymptotic
behaviour found in Refs.\cite{BK,PW} may depend on the prescription
for the additional poles of the propagator in temporal axial gauge.
In Sect.~5, the manifestly gauge invariant
Polyakov loop correlation (PLC)
is used as a basis to determine the screening function. 
As has been most recently shown in Ref.~\cite{BN}, the large-distance
behaviour of the PLC is strongly sensitive to the existence of a
magnetic mass, and its introduction leads to exactly the same
value of the Debye screening mass as obtained from the gauge-independent
pole of the gluon propagator.
In Sect.~6, next-to-leading order screening is discussed for
the scenario of a static $A_0$ condensate. While perturbation
theory involves even more uncertainties in this case, it indicates
a stronger sensitivity of the electrostatic
potential to the magnetic scale. In Sect.~7 recent
lattice results on nonabelian Debye screening are reviewed
which seem to favour enhanced exponential screening at
moderate distances.
The data are found to agree well with the results for the PLC
in the symmetric phase 
when a magnetic mass $m_{\mathrm m}\approx g^2T/4$ is included.

\section{A 
propagator-based definition of the Debye mass}

In linear response theory,
the chromoelectric field $\langle E^i(x)\rangle$ induced by a single
external source $J$ is fully determined by the gluon propagator,
without the need to consider higher vertex functions. With only one
source there is also only one direction in colour space so that
even the nonabelian field strength operator is linear in the
gauge potentials (the commutator terms vanish trivially), and also
covariant conservation of the external current,
$\langle D_\mu(A)\rangle
J^\mu=0$ reduces to simple transversality, $\partial_\mu J^\mu=0$.
In particular, the longitudinal electric field in momentum space
is given by\cite{Weldon}
\begin{equation}
\langle \tilde E^i_L \rangle = -i\frac{k^i}{{\bf k}^2}
\frac{K^2}{K^2-\Pi_L} \tilde J^0,
\label{El}
\end{equation}
where
\begin{equation}
\Pi_L=\Pi_{00}-\frac{k^ik^j}{{\bf k}^2}\Pi_{ij}-\frac{K^\mu K^\nu}{K^2}
\Pi_{\mu\nu},
\label{PiL}
\end{equation}
with $\Pi_{\mu\nu}$ the gluon self-energy
which is diagonal in colour space in the absence of
gauge symmetry breaking. (We follow the conventions of Ref.
\cite{Weldon} and denote 4-vectors by capital letters.)

In the static case, Eq.~(\ref{PiL}) reduces to $\Pi_L=\Pi_{00}$ and
the electric field induced by a static source $J^\mu(x)=Q\delta^\mu_0
\delta^3({\bf x})$ is determined by
\begin{equation}
\langle E^i({\bf x}) \rangle = \frac{\partial}{\partial x_i} \Phi(|{\bf x}|)
\end{equation}
with the potential $\Phi$
\begin{eqnarray}
&&\Phi(r)= Q\int\frac{d^3k}{(2\pi)^3}
\frac{e^{i{\bf k}{\bf r}}}{k^2+\Pi_{00}(k_0=0,k)}\nonumber\\
&=&\frac{Q}{(2\pi)^2}\int_{-\infty}^\infty
\frac{e^{ikr}-e^{-ikr}}{2ir} \frac{k\,dk}{k^2+\Pi_{00}(0,|k|)}.
\label{v}\end{eqnarray}

If $\Pi_{00}(0,k)$ is an even function in $k$, the last
two pieces in Eq.~(\ref{v}) can be evaluated by
closing the contour in the upper and lower half plane, respectively.
At leading order in the high-temperature expansion,
$\Pi_{00}(0,k)$ is just a constant,
\begin{equation}
\Pi_{00}(0,k)\Big|_{g^2T^2}\equiv m_0^2=\frac{e^2T^2}3
\label{m0}
\end{equation}
with $e^2=(N+N_f/2)g^2$ for colour group SU($N$) and $N_f$ flavors,
and Eq.~(\ref{v}) involves just simple poles at $k=\pm i m_0$,
yielding
\begin{equation}
\Phi(r)=\frac{Q}{4\pi r}e^{-m_0r}.
\label{Phi0}
\end{equation}

The next-to-leading order contribution to $\Pi_{\mu\nu}$ is down by
one power of $g$ rather than $g^2$ because of the
``plasmon effect''\cite{Kapusta}.
Whereas in Abelian theories $\Pi_{\mu\nu}$ is a
manifestly gauge independent
object, in the nonabelian case it will generally
depend on the gauge fixing parameters. Indeed, the infrared limit of
$\Pi_{00}$ turns out to be gauge dependent at the order $gm^2_0$,
viz. \cite{T}
\begin{equation}
\Pi_{00}(0,k\to0)=m^2_0\left(
1+\alpha {\sqrt{3N'}\over 4\pi} g \right)
\label{toi}\end{equation}
where $\alpha$ is the gauge parameter of covariant gauges and
\begin{equation}
N'\equiv {N\over 1+N_f/(2N)}.
\label{Np}
\end{equation}

However, this general gauge dependence does not mean that the screening
function $\Phi(r)$ as defined in Eq.~(\ref{v}) is altogether
unphysical.
If $(k^2+\Pi_{00}(0,k))^{-1}$ still has simple poles at $k=\pm i m$ one
can prove on an algebraic level that their position, and therefore $m$,
is gauge fixing independent\cite{KKR}. Hence, the
screening mass $m$ in the exponent of Eq.~(\ref{Phi0}) can be
a physical quantity, while the pre-exponential factor
will depend on the gauge choice. This leads to the self-consistent
determination of the Debye mass through\cite{AKR}
\begin{equation}
m^2 = \Pi_{00}(0,k)\Big|_{k^2=-m^2}
\label{mdef}\end{equation}
rather than $\Pi_{00}(0,k\to0)$ which is usually taken as its
definition\cite{Kapusta}.

The identification of $\Pi_{00}(0,k\to0)$ with the screening mass is
in fact deficient already in the Abelian case.
The infrared limit of $\Pi_{00}$
is directly related to the second derivative of the
thermodynamic potential
with respect to the chemical potential\cite{Fradkin}, and from this
one knows\cite{Kapusta}
\begin{equation}
\Pi_{00}(0,k\to0)={e^2T^2\over 3}
\left(1-{3e^2\over 8\pi^2}+{\sqrt3 e^3\over 4\pi^3}
+\ldots\right).
\label{pi00qed}
\end{equation}
However, this result is not renormalization-group invariant. This is
repaired by adopting (\ref{mdef}), which amounts to adding
\begin{equation}
\Pi_{00}(0,k)\big|_{k^2=-m^2}
-\Pi_{00}(0,k\to0)=
{e^2T^2\over 3}\left({2e^2\over 9\pi^2}-{e^2\over 6\pi^2}
\left[\ln{\tilde\mu\over \pi T}+\gamma_E\right]
+O(e^4)\right),
\label{dmqed}\end{equation}
where $\tilde\mu$
is the mass scale introduced by dimensional regularization
in which minimal subtraction has been performed.
The coefficient of the
logarithmic term in (\ref{dmqed}) is exactly such that
$\partial  m^2/\partial  \tilde\mu=0$
because $de/d(\ln\tilde\mu)=\beta(e)=e^3/(12\pi^2)+O(e^5)$.

Turning again to the nonabelian case, it is now clear that we need
more than only the infrared limit (\ref{toi})
of the next-to-leading order gluon self-energy.
Since we are only considering the static case here, it is
in fact possible to give the complete next-to-leading order
result for $\Pi_{00}$. In general such a calculation would involve
the rather complicated propagators and vertices of the Braaten-Pisarski
resummation program. However, as shown in Refs.~\cite{AE,SED},
for just the next-to-leading order contribution in
static Green's functions, this resummation scheme boils down
to the simpler ring resummation of Gell-Mann and Brueckner\cite{GMB}.
There one has to keep only the zero-mode contributions in the
sum over Matsubara frequencies, and resummation consists
only of the inclusion of the Debye mass in the longitudinal gluon
propagators.

The static ring-resummed propagator in general covariant as well as
Coulomb gauge (with gauge parameter $\alpha$) reads
\begin{equation}
\Delta_{\mu\nu}\bigg|_{p_0=0}=\left[
\frac1{{\bf p}^2+m^2}
\delta^0_\mu \delta^0_\nu
+\frac1{{\bf p}^2}
\left(\eta_{\mu\nu}-\delta^0_\mu \delta^0_\nu +
\frac{P_\mu P_\nu}{{\bf p}^2} \right)
-\alpha \frac{P_\mu P_\nu}{({\bf p}^2)^2} \right]_{P_0=0}.
\label{stpr}\end{equation}
In these gauges,
the complete next-to-leading order contribution to $\Pi_{00}(0,k)$
is found as\cite{AKR}
\begin{eqnarray}
&&\delta\Pi_{00}(k_0=0,{\bf k})
=gm_0\sqrt{3N'}\int
\frac{d^{3-2\varepsilon}p}{(2\pi)^{3-2\varepsilon}}
\biggl\{\frac1{{\bf p}^2+m^2}+\frac1{{\bf p}^2} \nonumber\\
&&\qquad+\frac{2(m^2-{\bf k}^2)}{{\bf p}^2({\bf q}^2+m^2)}
+(\alpha-1) ({\bf k}^2+m^2)
\frac{{\bf p}^2+2{\bf p}{\bf k}}{{\bf p}^4({\bf q}^2+m^2)} \biggr\},
\label{pi00}\end{eqnarray}
where
${\bf q}={\bf p}+{\bf k}$. (Here dimensional regularization has
been used when separating the static modes from the sum over
Matsubara frequencies\cite{AE}; the limit $\varepsilon\to0$ gives a
regular expression because of the odd integration dimension.)

The new definition (\ref{mdef}) requires to evaluate at $k^2=-m^2$.
There the gauge dependent piece proportional to $\alpha$ vanishes
algebraically, before doing the integrations, but the integral
itself is linearly singular on the ``mass-shell'' $k^2=-m^2$. However,
introducing a small infrared cutoff and taking the limit $k^2\to
-m^2$ before lifting the cutoff removes this term
completely (cp. \cite{BKS,BKSC}).
This can be achieved in a gauge invariant way either
by dimensional regularization or by taking the symmetric limit of
the Higgs mechanism\cite{Sintra}.

The third term of the integrand in (\ref{pi00}) is logarithmically
singular as $k^2\to-m^2$. This singularity is caused exclusively
by the massless denominator in the spatially transverse part of
the gluon propagator (\ref{stpr}).
A magnetic screening mass $m_{\mathrm m}$
would screen this singularity, and because the latter is only
logarithmic, the coefficient of the corresponding logarithm is
unambiguously determined by (\ref{pi00}),
\begin{equation}
\delta\Pi_{00}(0,k)\Big|_{k^2\to-m^2_{}}\to
{g^2 N m_{} T\over 2\pi}\ln{m_{}\over m_{m}}
\label{pi00sing}\end{equation}
up to terms that are regular as $m_{m}\to0$.
Assuming that $m_{m}\sim gm_{}$, the next-to-leading
order contribution to $m_{}^2$ is found to be of order
$g\ln(1/g)$ rather than $g$,
\begin{equation}
{\delta    m_{}^2\over m_{0}^2}=
\frac{\sqrt{3N'}}{2\pi}\,g\ln\frac1g+O(g),
\label{dln}\end{equation}
which is positive, at least at weak coupling $g\ll1$, contrary to
expectation\cite{Kapusta}.

The sublogarithmic terms cannot be calculated completely, because
the presumed phenomenon of magnetic screening is
nonperturbative\cite{Linde,GPY}. However, in order to obtain an {\it
estimate} of the full $O(g)$ contribution, let us assume that a simple
replacement of $1/{\bf k}^2\to1/({\bf k}^2+m^2_{m})$
in the transverse part of the static propagator (\ref{stpr})
correctly summarizes the effects at $k\sim g^2T$. Then we
may go on to evaluate the remaining contributions
in (\ref{pi00}), which leads to\cite{AKR,Sintra}
\begin{equation}
\delta\equiv
{\delta m^2_{}\over m^2_{0}}
=\frac{\sqrt{3N'}g}{2\pi}\frac{m}{m_0}\left(
(1-\frac{m_{\mathrm m}^2}{4m^2})\ln\frac{2m+m_{\mathrm m}}{m_{m}}
-\frac12-\frac{m_{\mathrm m}}{2m}\right)+O(g^2).
\label{d}\end{equation}

{}From the point of view of the original Braaten-Pisarski resummation
programme\cite{BP}, the Debye mass $m$ appearing
on the right-hand side of the above equations as well as in
Eq.~(\ref{mdef}) should be identified with
the HTL value $m_0$ in order to be consequently perturbative.
However, by the very introduction of a magnetic mass term we have
already stepped beyond the resummation of HTL contributions.
One could therefore equally well consider a resummation of the
correction terms to the classical value of the Debye mass at
the dressed one-loop order, and determine $m$ from Eq.~(\ref{d}) in
a fully self-consistent manner. Diagrammatically, this corresponds
to including not only ``daisy'' but also ``super-daisy''
diagrams\cite{DJ}.

Of course, all this does not affect the
order $O(g)$ under consideration, but for realistic
values of $g$ the two possibilities will lead to noticeable
differences numerically. The functional form of Eq.~(\ref{d}) is
such that a self-consistent evaluation will generally give
larger corrections $\delta$.
We shall later adopt the variation caused by this as a measure of the
uncertainty of the one-loop result, together with the more conspicuous
uncertainty from the value of the hypothetical
magnetic mass $m_{\mathrm m}$.

\section{Chromoelectric screening functions in covariant gauges}

Let us now ignore for the moment that the full function $\Phi(r)$ is
gauge dependent and consider what screening behaviour it is purporting
beyond the presupposed Yukawa form.

Away from the singular points $k=\pm i m$, $\delta\Pi_{00}(0,k)$
as given by Eq.~(\ref{pi00}) is regular so that there seems to be
no need for the introduction of a magnetic mass as concerns the
evaluation of Eq.~(\ref{v}). Without a magnetic mass,
$\Pi_{00}(0,k)$ including the next-to-leading order is given by
\begin{equation}
\Pi_{00}(0,k)=m_0^2+\frac{g\sqrt{3N'}m_0m}{2\pi}\left[
\frac{m^2-k^2}{mk}\arctan\frac{k}{m}+\frac{\alpha-2}2 \right].
\label{pi00k}
\end{equation}
If this expression is used to compute the full screening
function $\Phi(r)$
from Eq.~(\ref{v}), one finds a surprising behaviour\cite{BK}:
$\Phi(r)$ decays exponentially, but the pre-exponential
factor oscillates before approaching a negative constant for very
large $r$.

This strange result can be understood by inspecting the
analytic structure of
$D_L^{-1}\equiv(k^2+\Pi_{00}(0,k))$. In the left half of
Fig.~1 this is displayed for
the first quadrant of the complex $k$ plane (the others are given by
reflection on the real and imaginary axes). At $k/m_0=i$,
$D_L^{-1}$ no longer
has a simple zero, but a logarithmic branch singularity, and there is
a branch cut from $i$ to $\infty$. The original
zero of $D_L^{-1}$, however,
still exists: it has moved to the right (and also left)
of the imaginary
axis. These complex poles of $D_L$ lead to an oscillatory behaviour of
$\Phi(r)$; for the particular coupling constant chosen
in Fig.~1, $t\equiv
g\sqrt{3N'}/8=0.25$, and gauge parameter $\alpha=1$,
they contribute a term proportional to
$\cos(0.313x)e^{-1.208x}/x$, where $x\equiv r m_0$. There is, however,
also the contribution from the cut, which adds a term $-f(x)e^{-x}/x$
with a strictly positive function $f$, so that asymptotically, for
very large $x$, the behaviour is that of a repulsive Yukawa potential
with screening mass $m_0$. Both, the function $f$ and the position of
the complex pole, are gauge dependent.

\begin{figure}                                                   
\centerline{ \epsfxsize=6.5in \epsfbox{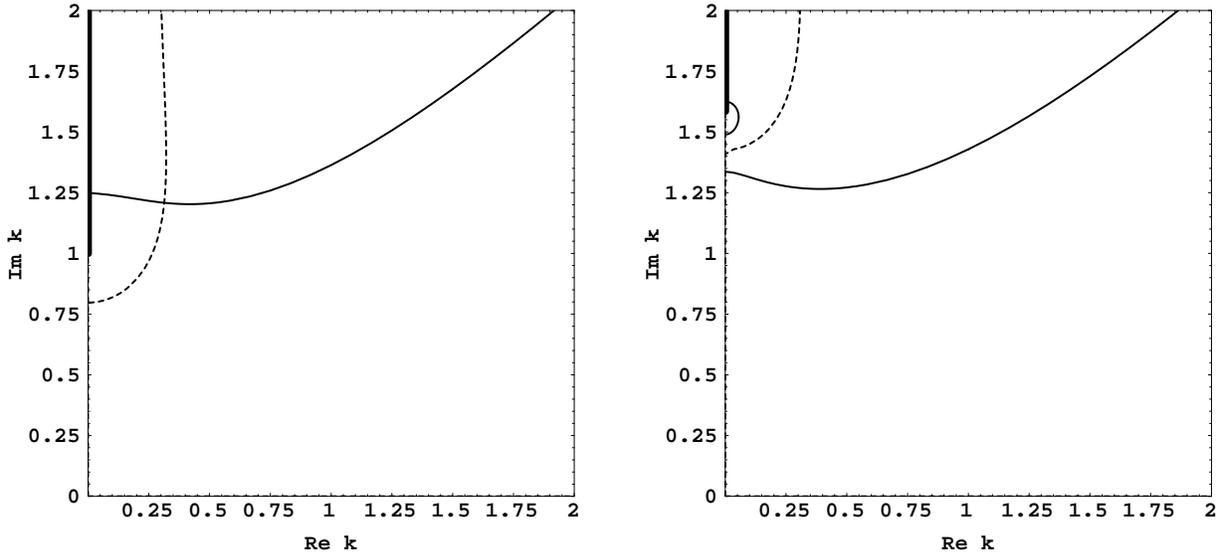} }
\caption[f1]{\label{f1}{The analytic structure of $D_L^{-1}\equiv
k^2+\Pi_{00}(0,k)$ in units where $1=m_0$ and with coupling
$t\equiv g\sqrt{3N'}/8=0.25$. The left side corresponds to
the case of zero magnetic mass, the right side to $m_{\mathrm m}/m=0.25$.
Full lines correspond to $\Re e D_L^{-1}=0$, dashed ones
to $\Im m D_L^{-1}=0$, and their intersections to poles of
$D_L$. The thick part of the imaginary axis marks the location
of the branch cut; otherwise the imaginary part vanishes on both the
real and imaginary axes. On the left, there is one (gauge dependent)
complex pole; on the right, there are two poles on the imaginary axis:
the lower one is at the gauge-independent position $k=im$, and there
is a gauge-dependent one just below the branch singularity. }}   
\end{figure}                                                     

However, if one allows for a small magnetic mass $m_{\mathrm m}$,
the analytic structure,
and therefore $\Phi(r)$, changes drastically. $\Pi_{00}$ then becomes
\begin{eqnarray}
\Pi_{00}(0,k)&=&m_0^2+\frac{g\sqrt{3N'}m_0}{2\pi}
\biggl[-{1\over2} m-{1\over2} m_{\mathrm m}\\ \nonumber
&&+{1\over k}(m^2-{1\over2} m_{\mathrm m}^2-k^2)
\arctan{k\over m+m_{\mathrm m}} \\ \nonumber
&&+\left(k^2+m^2\right)\Bigl\{ {k^2+m^2\over m^2_m k}
\Bigl(\arctan{k\over m+\sqrt\alpha m_{\mathrm m}}\\ \nonumber
&&\qquad -\arctan{k\over m+m_{\mathrm m}}\Bigr)
+(\sqrt\alpha-1){1\over m_{\mathrm m}} \Bigr\}\biggr],
\end{eqnarray}
where $m_{\mathrm m}$ has been introduced in a gauge-invariant manner by
mimicking the Higgs mechanism\cite{Sintra}.
Choosing $m_{\mathrm m}/m=t=0.25$,
Eq.~(\ref{mdef})
gives a self-consistent Debye mass of $m=1.335m_0$, and for these
parameters the analytic structure of $D_L^{-1}$ is rendered in the
right half of Fig.~1.
There is now a simple zero on the imaginary axis at $k=im$, and
the logarithmic branch singularity has moved further up to
$k=i(m+m_{\mathrm m})$.
The lines $\Re e D_L=0$ and $\Im m D_L=0$ no longer intersect at
complex values of $k$, but only on the imaginary axis. There is also
a gauge dependent zero very close to the branch singularity, but the
dominant singularity in $D_L$ is the one at $k=im$ with $m$ gauge
independent. Thus $\Phi(r)$ no longer oscillates but decays
exponentially
for large $x$ with gauge independent screening mass $m$.

The strong dependence of the analytic structure of $D_L$ on the
infrared behaviour of the transverse gluons of course means that
perturbation theory cannot be trusted. Close to the branch singularity
of $D_L^{-1}$, which arises at next-to-leading order, the correction
terms become larger than the leading-order ones, and one clearly has to
expect even more important contributions from higher orders.
The appearance of gauge dependent poles is therefore just a
manifestation of the incompleteness of the results in the vicinity
of $k=\pm i m_0$. Assuming that all infrared singularities
cure themselves in the complete
results, e.g.~by the generation of a magnetic mass,
all gauge dependent singularities have to disappear according to
the gauge dependence identities of Ref.~\cite{KKR}.
Whether in the final result the pole of the leading-order
result survives
at a shifted location cannot be established within the present
resummed perturbation theory. However if it indeed does,
Eq.~(\ref{dln})
gives the leading order contribution to such a shift; the
sublogarithmic terms of Eq.~(\ref{d}) on the other hand are just
a simple-minded estimate.

As we have seen, defining a chromoelectric screening function on the
basis of the gauge-dependent gluon propagator allows one at most to
extract a gauge independent exponential screening behaviour. All other
details have to be considered unphysical. One longstanding proposal for
doing better is to use a particular ``physical'' gauge\cite{Kapusta},
to wit, the temporal axial gauge.

\section{Non-Debye screening in temporal axial gauge?}

In recent work by Baier and Kalashnikov\cite{BK} and, more rigorously
following the Braaten-Pisarski resummation program, by Peign\'e and
Wong\cite{PW}, the screening function $\Phi(r)$ has been evaluated
in temporal axial gauge (TAG), $A_0=0$, and an asymptotic behaviour has
been reported which differs qualitatively from the one obtained in
covariant gauges. At very large distances $\Phi(r)\sim -1/r^6$ has
been found, i.e.~repulsive power-law behaviour, after an oscillatory
regime at intermediate distances.

The resummed calculation in TAG differs from the corresponding one
in other gauges in that one cannot at once restrict to the zero modes
of the resummed one-loop result. In TAG, the gluon propagator contains
poles $1/p_0^2$, which bring in contributions from resummed vertices,
which would otherwise simply vanish in the static limit.
These additional
terms in $\Pi_{00}(0,k)$ turn out to involve odd powers of $k$, i.e.
a branch point at the origin of the $k^2$ plane,
and this is responsible
for the asymptotic power-law behaviour.

As mentioned in the introduction, a power-law asymptotic
behaviour is known also from
higher-order contributions to screening in a non-relativistic
electron gas\cite{CM}. More surprising here is the reversed sign,
and also that such a behaviour should arise already at the
next-to-leading
order, while the results in covariant and Coulomb gauges have a
well-behaved
Taylor series expansion in $k^2$. 

Let us recall that the motivation for employing the rather awkward
TAG is
that there the electric field strength
is linearly related to the vector potentials: $E_i=-\partial_0 A_i$.
The longitudinal gluon
propagator $D_L$, when computed in this gauge,
is thus directly related to the correlation of two
$E$ operators.

It has been argued\cite{KK,HKT,Kapusta} that the static
correlation $\langle
E_L({\bf x}) E_L(0) \rangle$ is furthermore directly related to the
free energy of a separated quark-antiquark pair in the
singlet state and thus physically measurable.
However, in this reasoning one has to consider external sources
$\cal E$
which couple to the electric field strength operator
in order to construct a perturbing Hamiltonian $H^{\rm ext}=
\int d^3x E_j^a({\bf x},t) {\cal E}_j^a({\bf x},t)$. But $E_j^a$ is
not a gauge invariant operator and the correlation of two $E$ operators
at separate points thus may depend on the gauge fixing parameters.
Indeed, one finds that under a change of gauge condition
$f_\mu A^\mu\to
(f_\mu+\delta    f_\mu)A^\mu$ the correlation of two
chromoelectric field
operators varies according to\cite{KKRS}
\begin{equation}
\delta\langle E^a_j(x)E^e_k(y)\rangle =-gf^{abc}\int d^4\!z
\langle  E^b_j(x)\bar  c^c(x)c^d(z)\delta
f_\mu A^{d\mu}(z)E^e_k(y)\rangle
+(a,j,x \leftrightarrow e,k,y),
\end{equation}
where $\bar  c$ and $c$ are Faddeev-Popov ghost fields, which make
their appearance even in gauges which are otherwise ghost-free.

\begin{figure}                                                   
\centerline{ \epsfxsize=5in \epsfbox{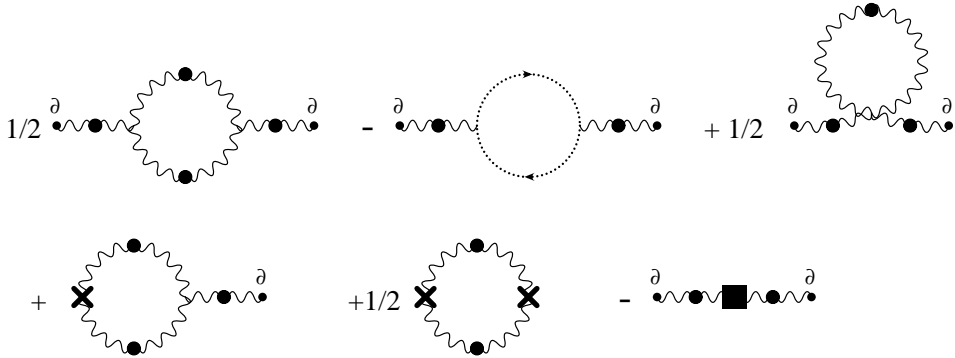} }
\caption[f2]{\label{f2}{One loop diagrams contributing to the
correlation
of two chromoelectric field strength operators.
The wavy line with a dot is the ring-resummed
gluon propagator and the dotted line is the Faddeev-Popov propagator.
The crosses denote the extra vertices from the commutator term in
the nonabelian field strength operator and the small box refers to
the insertion of the Debye mass as a counterterm of the resummed
perturbation theory.}}       
\end{figure}                                                     

At the expense of having to keep the nonlinear terms in $E$, one can
without further difficulty perform a resummed calculation of $\langle
E_L(-k)E_L(k)\rangle$ in gauges other than TAG. In order to obtain
the next-to-leading order term, in almost any gauge except TAG one
can restrict oneself to the zero-mode contributions, which more
than repays the complications from the nonlinearity of $E$. A gauge
which like TAG has a straightforward Hamiltonian formulation is the
Coulomb gauge, which has the same zero-mode propagators as the
covariant
gauge.
Evaluating the diagrams shown in Fig.~2 in Coulomb gauges with
gauge parameter $\alpha$, one obtains
\begin{eqnarray}
\langle E_L^a(-k)E_L^b(k)\rangle &=& {i\delta^{ab}\over k^2+m^2}
\biggl\{
1-\\ \nonumber
&&\quad-{g^2NmT\over (2\pi)(k^2+m^2)} \Bigl[ {m^2-k^2\over mk}
\arctan{k\over m}
-1+{\alpha\over 2} \Bigr] \\ \nonumber&&\quad
+{g^2NmT\over 2\pi k^2}\Bigl[ {m^2-k^2\over mk}\arctan{k\over m}
-1\\ \nonumber&&\qquad\qquad
+{\alpha\over 2}\left(1-{k^2+m^2\over mk}\arctan{k\over m}\right)
\Bigr] \\ \nonumber&&\quad
+{g^2NmT(k^2+m^2)\over 8\pi k^4}\Bigl[ {k^2-m^2\over mk}
\arctan{k\over m}
+1\\ \nonumber&&\qquad\qquad
+\alpha\left(1-{k^2+m^2\over mk}\arctan{k\over m}\right)\Bigr]
\biggr\}.
\label{EE}\end{eqnarray}
The gauge parameter $\alpha$ does not drop out from this
expression, which
clearly shows that $\langle
E_L({\bf x}) E_L(0) \rangle$ cannot be a physical quantity. It thus cannot
be directly related to the free energy of two separated quarks,
contrary
to what is assumed in Ref.~\cite{KK,HKT,Kapusta}.

In view of the non-Debye screening behaviour found
in TAG, it is interesting to note that in covariant and
Coulomb gauges $\langle
E_L({\bf x}) E_L(0) \rangle$ again has a series expansion in powers
of $k^2$
rather than $k$. This suggests that the different analytic
behaviour found
in TAG may well be an artefact of the treatment of
the  $1/p_0^2$ in the gluon propagator.
In fact, such poles remain in the final expressions derived in
Refs.~\cite{BK,PW}, and are then evaluated by a principal-value
prescription.

But already at zero temperature, the principal-value prescription
in axial
gauges has been shown to be flawed in Wilson-loop
calculations\cite{CCM},
and at finite temperature the situation is even worse. $A_0=0$ is
incompatible with strict periodicity in imaginary time\cite{GPY},
so when
using the imaginary-time formalism (which is convenient for
separating the zero-mode contributions) one has either to give up
periodicity in the longitudinal propagator, which leads to rather
complicated Feynman rules\cite{KPVL}, or one has to relax the condition
$A_0=0$. A regularisation of TAG through general axial gauges has been
developed by Nachbagauer\cite{HN}, and this procedure would again
eliminate contributions from the resummed vertices, thus presumably
removing the source of the contributions giving odd powers of $k$.
And in the absence of odd powers of $k$, one would fall back to the
task of finding the shift of the pole position, if any. If one defines
an effective $\Pi_L$ from the electric correlation
(cf.\ Ref.\cite{HKT})
as obtained in Coulomb or covariant gauges, Eq.~(\ref{EE}),
all the additional contributions are suppressed by powers of
$(k^2+m^2)$,
so that indeed the same Debye mass is obtained from Eq.~(\ref{mdef})
as through the conventional gluon self-energy.

{}From all that it seems unlikely that a nonexponential asymptotic
behaviour
of Debye screening should occur already at relative order $g$. It
might of course arise at higher orders, and in view of the largeness
of $g$ in actual QCD this would probably be an important effect
already on not so large distances.

\section{Gauge invariant screening from a Polyakov loop correlation}

We have seen that the chromoelectric screening functions defined from
the gluon propagator as well as from
the correlation of electric field strength operators
are strongly gauge dependent in general and that only the exponential
decay contributed by a singularity
of the gluon propagator will be gauge independent and therefore
of physical significance.
If one is interested in the detailed screening behaviour beyond the
value of
the supposed electric screening mass, it is mandatory to
first find a manifestly gauge invariant
definition.

A natural choice which can also be implemented
without difficulty in lattice gauge theory
is the Polyakov loop correlation (PLC) \cite{N1}. The Polyakov
loop operator
at the spatial point ${\bf x}$ is defined by
\begin{equation}
\Omega({\bf x})=\frac1N P\exp \left(-ig\int_0^\beta d\tau A_4({\bf x},\tau)
\right)
\end{equation}
where $P$ denotes path ordering and $\tau$ is the imaginary time.
The correlation of two Polyakov loop operators is directly related to
the free energy of a quark-antiquark pair at the same
separation\cite{McLS}.
At lowest order (resummed tree-level), the connected part
is given by
\begin{equation}
\langle {\rm tr} \Omega^\dagger (r) {\rm tr} \Omega(0) \rangle_c-1=
\frac{N^2-1}{8N^2} \left( \frac{g^2e^{-m_0r}}{4\pi rT} \right)^2
+ O(g^6)=
f^{(0)}_{\rm m.s.}+O(g^6)
\label{ms0}
\end{equation}
and is given by diagram (a) of Fig.~3.

\begin{figure}                                                   
\centerline{ \epsfxsize=5in \epsfbox{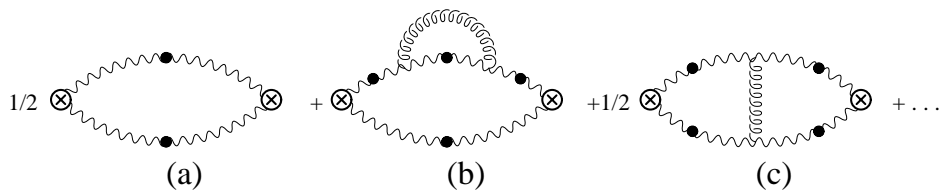} }
\caption[f3]{\label{f3}{Resummed tree-level (a) and one-loop (b,c)
diagrams contributing to the mean-square part of the Polyakov loop
correlation. Here the wavy lines with a dot represent
static longitudinal
propagators while the transverse ones have been drawn as springs.
Diagrams with vanishing next-to-leading order contribution
have been omitted.
Subtraction of the Debye mass counterterm
is understood to be included in diagram (b).
With a finite magnetic mass, there is also a seagull-type
self-energy insertion to be added to (b).}}       
\end{figure}                                                     

The one-loop resummed correction
to the mean-square correlation, $f^{(1)}_{\rm m.s.}$ is given by the
remaining diagrams of Fig.~3.
Inspecting first large $x\equiv m_0r$, one obtains\cite{N1}
\begin{eqnarray}
\label{ff}
&&f^{(0)+(1)}_{\rm m.s.}/f^{(0)}_{\rm m.s.} \equiv F(x) \\ \nonumber
&=&1+\frac{\sqrt{3N'}g}{2\pi}
\left\{ x(-\frac12\ln x-\ln2-\frac\gamma2+\frac32)+O(\ln x) \right\}
\end{eqnarray}
with $\gamma$ being Euler's constant and $N'$ as defined in
Eq.~(\ref{Np}).
The correction terms that are linear in $x$ could be absorbed
by a shift
of the lowest order Debye mass $m_0^2\to m_0^2+\delta m^2$, but there
is also a term proportional to $x \ln x$, which, if exponentiated,
would
lead to a potential that falls off faster than any Yukawa potential
squared.
Closer inspection of the Feynman diagrams\cite{BN}
reveals, however, that the 
contributions $\propto x \ln x$ arise from singularities in momentum
space which are strongly sensitive
to the magnetic mass scale.

Indeed, introducing again a small magnetic mass $m_{\mathrm m}$
(see the appendix for details)
changes the result (\ref{ff})
qualitatively
for $x\gg m/m_{\mathrm m}$. The term $\propto x \ln x$ contributed by
diagram 3b arises from a branch singularity coinciding
with a double pole at ${\bf k}^2=-m^2$. A magnetic mass $m_{\mathrm m}$
separates these singularities and instead leads to a term
linear in $x$. Its coefficient is given by the next-to-leading
order correction to the Debye mass as derived from the
gauge-independent pole of the gluon propagator in Sect.~2,
\begin{equation}
F^{(b)}(x)=
\frac{\sqrt{3N'}g}{2\pi} \left\{
-x\left[(1-\frac{m_{\mathrm m}^2}{4m^2})\ln\frac{2m+m_{\mathrm m}}{m_{m}}
-\frac12-\frac{m_{\mathrm m}}{2m} \right]
+O(x^0) \right\}
\label{f3b}
\end{equation}

The term $\propto x \ln x$ contributed by
diagram 3c comes from a simple pole coinciding with a branch
singularity at ${\bf k}^2=-4m^2$. A small magnetic
mass replaces the pole by
two neighbouring branch singularities\cite{BN}, yielding
\begin{equation}
F^{(c)}(x)=
\frac{\sqrt{3N'}g}{2\pi} \left\{
\frac{m^2}{m_{\mathrm m}^2}(\ln\frac{2m_{\mathrm m}x}{m}+\gamma)
+O(m/m_{\mathrm m})
\right\},
\label{f3c}
\end{equation}
which grows only logarithmically at large $x$.

In these results, the unphysical gauge modes have been made massive, too.
If they were kept massless, they would contribute gauge-parameter
dependent terms proportional to $x$ to both $F^{(b)}$ and $F^{(c)}$
with opposite sign, leaving their sum unchanged.
Absorbing all the linear terms of $F(x)$ by a shift of the
Debye mass implies
$F(x)\to F(x)+x{\delta m^2}/(mm_0)$,
thus leads to exactly the same next-to-leading order result
for the Debye mass as from the gauge-independent
pole of the gluon propagator,
Eq.~(\ref{d}).

In order to inspect the PLC screening function in more detail
we shall need the form of $F(x)$ also for moderate values of $x$.
For $x\ll m/m_{\mathrm m}$, the complete
correction to the
pre-exponential screening function $F(x)$ of Eq.~(\ref{ff})
reads\cite{N1}
\begin{eqnarray}
F(x)&=&1+\frac{\sqrt{3N'}g}{2\pi} x\biggl\{
-\frac12\ln x-\ln2-\frac\gamma2+\frac32 \\ \nonumber &&
-\frac12 g(2x)+\frac{\ln (2x)+\gamma}x+\frac12 e^{4x}{\rm Ei}(-4x)
-\frac1x e^{2x}{\rm Ei}(-2x)
\biggr\}
\label{SN}
\end{eqnarray}
where Ei($x$) is the exponential integral\cite{GR}, $x\equiv m_0r$, and
\begin{equation}
g(x)=\int_0^\infty \frac{dz}{z+1}e^{-xz}\ln\frac{z+2}z. 
\label{gx}\end{equation}
For large $x$, $g(x)\sim \ln(x)/x.$

Including the magnetic mass $m_{\mathrm m}$ makes a difference
for $x\gtrsim m/m_{\mathrm m}$
and 
we instead obtain
\begin{eqnarray}
F(x)&=&1+
\frac{\sqrt{3N'}g}{2\pi} \frac{m_0}{m}
\biggl\{
\ell(\bar x)-\frac{\bar x}2 g(2\bar x)
+\bar x h(2\bar x)+\frac{\bar x}2 h(4\bar x) \\ \nonumber
&&+\frac{\mu^2}4 \ln\frac{2+\mu}{\mu}
+(1-\frac{\mu^2}4)(1-e^{-\mu \bar x})
\left(\frac1{\mu}+\frac1{2+\mu}\right)
\\ \nonumber
&&+e^{-\mu \bar x} \biggl[ -(1-\frac{\mu^2}4) \bar x
\left\{h(\mu \bar x)+h((2+\mu)\bar x)\right\}
\\ \nonumber
&&\qquad +(1+\frac{\mu^2}4)\left\{
h(\mu \bar x)-h((2+\mu)\bar x)\right\}
+\ln\frac{2+\mu}{\mu} 
\biggr]
\biggr\},
\label{fmx}\end{eqnarray}
where $\bar x=m r=xm/m_0$ and $\mu=m_{\mathrm m}/m$.
We have further introduced
$$h(x)\equiv e^x {\rm Ei}(-x)$$
and the function
\begin{equation}
\ell(x)\sim\cases{{x\over2}(\ln(x)+\gamma) & for $ x \ll 1/\mu $\cr
               \mu^{-2} \ln(x) & for $ x \gg 1/\mu $\cr}
\end{equation}
which we approximate by
\begin{equation}
\ell(x)\approx \frac{x}{2+\mu^2 x}(\ln(x)+\gamma).
\label{ell}\end{equation}


In Fig.~4 the screening
function $F(x)\exp(-2xm/m_0)$ is plotted logarithmically
for the cases without and with a magnetic mass for the parameters
used in Sect.~2.
In order to get a notion of the uncertainties of the perturbative
calculations, the next-to-leading order Debye mass $m$ is evaluated
both fully self-consistently (solving the nonlinear equation
defined by Eq.~(\ref{d})) and strictly perturbatively (replacing
$m\to m_0$ in all quantities that are already of relative order $g$).
One finds that for large $x$ the dominant feature is exponential decay
determined by the value of $m$. At small up to moderate distances the
pre-exponential
function is almost equally important. Notice that the
effective Debye mass given by the slope of the
tangent to the curves of Fig.~4 is
generally smaller than the asymptotic value.
\begin{figure}                                                   
\centerline{ \epsfxsize=5in \epsfbox{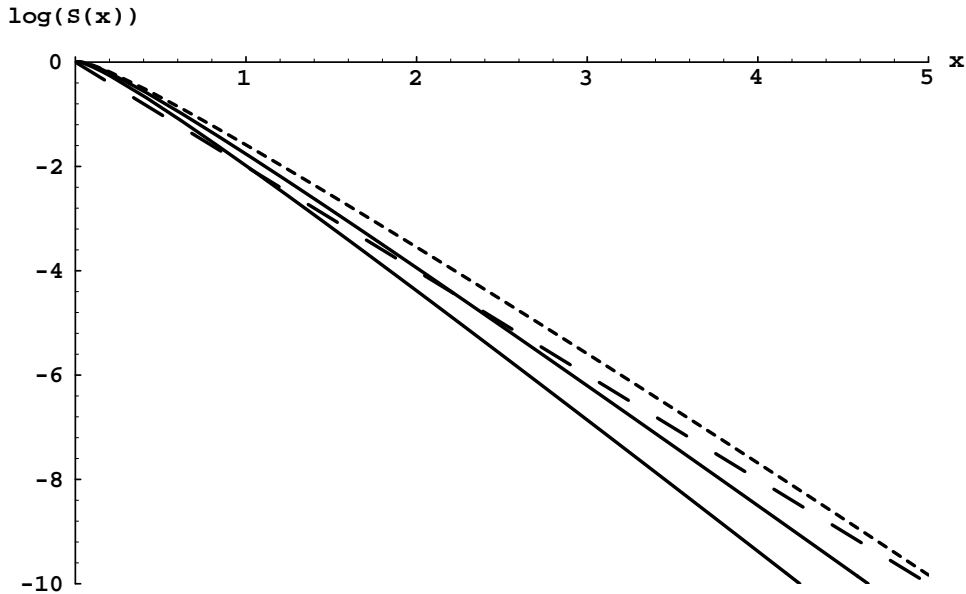} }
\caption[f4]{\label{f4}{The logarithm of the
screening function $S_{\rm PLC}=
F(x)\exp(-2xm/m_0)$ of the Polyakov loop correlation
for $t\equiv g\sqrt{3N'}/8=0.25$,
both without and with a magnetic mass. The short-dashed curve gives the
case without a magnetic mass and $m=m_0$,
the full lines correspond to $m_{\mathrm m}/m_0=0.25$.
The upper full line
is for $m$ from a fully self-consistent evaluation of Eq.~(\ref{d}),
the lower one for a strictly perturbative one as discussed in the text.
The long-dashed line marks the classical result, $\ln S=-2x$.
 }}       
\end{figure}                                                     

For the range of $x$ covered by Fig.~4, the dependence of the
screening function on the value of the magnetic mass is predominantly
through the next-to-leading order value of the Debye mass rather
than the explicit dependence of $F(x)$ on $m_{\mathrm m}$.
The latter dependence becomes relevant only for $x\gg 1/\mu$, but
there the perturbative result breaks down at some point, for
$F(x)$ also grows $\gg 1$.

\section{Screening with $A_0$ condensate}

It has been speculated\cite{A0,N2} that the infrared divergences
in the static
sector of high-temperature nonabelian gauge theories cause the
zero mode of $A_0$ to develop a vacuum expectation value which
breaks SU($N$) down to U(1)$^{N-1}$. Nadkarni\cite{N2} has considered
the effects of this scenario on chromoelectric screening in
ring-resummed perturbation theory. From calculations
in a unitary (diagonal) gauge,
he suspected a strong modification of the Debye mass.

In the diagonal gauge, where $A_0$
only involves the diagonal generators of SU(N),
correlation functions of the former correspond to the
gauge invariant correlations of the eigenvalues of the Polyakov loop
operator.
The zero mode of  $A_0$ is restricted to
$\hat v+\epsilon^{\hat a}T^{\hat a}$ where $T^{\hat a}$ are the
diagonal generators and $\hat v$ is the diagonal vacuum
expectation value.
In this parametrization, a nonvanishing $\hat v$ gives mass to
the off-diagonal
transverse static gluons $\tilde A$,
which couple to the electrostatic fluctuations
$\hat\epsilon$ through
$${\rm tr}[\tilde A,\hat v+\hat \epsilon]^2,$$
but there is no coupling between two electrostatic potential with
only one transverse gluon. Since this was the only vertex in the
next-to-leading order calculations in covariant and Coulomb gauges,
it was maintained in Ref.~\cite{N2} that all the results
$\Pi_{00}\propto gm^2$ were
pure gauge artefacts. However, by inspecting the perturbative series
in diagonal gauge, one finds that it ceases to exist for $\hat v\to0$.
Denoting the mass acquired by the off-diagonal transverse static fields
by $\mu=c g^2 T$, where the value of $c$ is not determined by
perturbation theory, the loop expansion parameter turns out to be
$1/c$, which diverges in the symmetric limit. Hence, no statement
can be made for the latter.

Nevertheless, with $c$ sufficiently large, the perturbation theory in
diagonal gauge again
makes sense and is in fact rather different from
the case of no $A_0$ condensate.
The correlation of
two $\hat\epsilon$ operators is manifestly gauge invariant
by virtue of its relation to the Polyakov loop eigenvalues.
In Ref.~\cite{N2} this correlation was considered at resummed one-loop
order only in the limit $k^2\ll \mu^2$. With the Feynman rules derived
in Ref.~\cite{N2} it is straightforward to obtain the complete
expression, which reads
\begin{equation}
\Pi_{00}(0,k)=m_0^2-\frac{\mu^2}{c\pi} \biggl\{ 1+
\left(2+\frac{k^2}{\mu^2}+\frac{k^4}{4\mu^4} \right)
\frac\mu{k}\arctan\frac{k}{2\mu}+\frac{k^2}{2\mu^2} \biggr\}.
\label{n2}\end{equation}
For small $k$ this coincides with the result given in Ref.~\cite{N2}.

The analytic structure of Eq.~(\ref{n2}) is quite different from
the ones we have encountered before.
Because the correction term in Eq.~(\ref{n2}) is negative for large $k$
and grows $\sim k^3$, there is generally a pole of
$(k^2+\Pi_{00}(0,k))^{-1}$
on the {\em real} $k$-axis. This is of course the dominant singularity
for large distances and, taking the principal value, it
would imply an oscillatory behaviour $\sim \cos(\lambda r)/r$. In fact,
the residue is such that it even may start out repulsive at small
distance,
as shown in Fig.~5. However, this phenomenon is linked with the
behaviour of the correction term in Eq.~(\ref{n2})
at large values of $k/\mu$, where it
blows up due to the unfavourable
large-momentum behaviour of the gluon propagator in the
unitary diagonal gauge. Judging from Eq.~(\ref{n2}), the new
perturbation series breaks down for $(k/\mu)^3\gtrsim c(m/\mu)^2$,
which is
in fact the region where the pole on the real axis arises. Since the
other contributions to the screening function are
suppressed at large $k/\mu$, it seems reasonable to just exclude the
contribution of the pole on the real axis while keeping the other
analytic structures at smaller $k$. This
can easily be done by a different choice of the integration contour
in Eq.~(\ref{v}).

\begin{figure}                                                   
\centerline{ \epsfxsize=5in \epsfbox{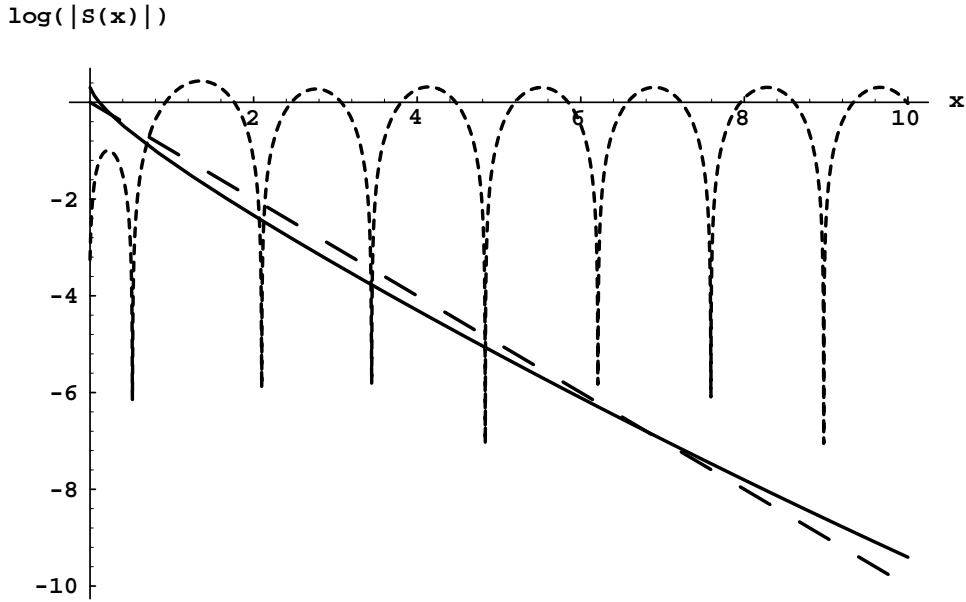} }
\caption[f5]{\label{f5}{The logarithm of the screening function $S(x)$
of the Polyakov loop eigenvalue correlation
in the case of a nonzero $A_0$ condensate with $c=1$ and
$\mu/m_0=0.25$, where $S(x)$ is normalized
such that $S_{\rm cl.}=\exp(-x)$. The short-dashed curve gives
the oscillatory behaviour when the pole on the real axis is
included (see text), which starts out with a reversed (repulsive)
sign. The full curve is the result upon subtraction
of the contribution from the real pole,
and the long-dashed line marks the the classical
screening function.}}       
\end{figure}                                                     

With $\mu > m_0/2$ the only singularity is then the one at
$k\approx\pm i m_0$,
so the asymptotic behaviour would be the classical one with only
small corrections. However
for $\mu<m_0/2$, which should be fulfilled at least with sufficiently
small $g$, there is a branch
cut on the imaginary axis superseding the original pole at
$k=\pm im_0$.
Hence, in this case,
the magnetic mass scale is the dominant one in the asymptotic
behaviour of the electric correlations. However, the imaginary part of
$(k^2+\Pi_{00}(0,k))^{-1}$ along the cut
is strongly peaked at $k\approx\pm i m_0$,
so this dominance of the magnetic scale
will occur only at very large distances.
At intermediate distances the effective Debye mass is smaller than
but close to the
classical one, as shown in Fig.~5 by the plot where the real
pole contribution
has been subtracted.

Altogether the resummed one-loop result for Debye screening
with an $A_0$ condensate appears more uncertain than with the
conventional vacuum. The entire correction term now is proportional
to the coefficient $c$ of the magnetic mass of the off-diagonal
transverse
gluons\footnote{The magnetic mass
of the diagonal gluons does not enter here and is
in fact independent of the one provided by the assumed dynamical
symmetry breaking.}. The value of $c$ is inherently nonperturbative
like the value of the magnetic mass which was invoked in the previous
sections, but there the dependence was only logarithmic.
A large value of $c$ would provide a small loop expansion
parameter of the new perturbation series
obtained in the diagonal gauge, at least for sufficiently small
momenta,
but one can only expect $c\sim1$.

\section{Comparison with lattice simulations}

Recently, high-precision lattice simulations of the Polyakov loop
correlation in pure SU(2) gauge theory have been performed by Irb\"ack
et al.\cite{ILMPR}. These simulations
employ temperatures up to nearly 8 times the critical
temperature so that one may hope that perturbation theory
becomes applicable.
Moreover, in SU(2) there exist also some (rather old)
lattice results
on the value of the hypothetical magnetic screening mass, so 
we can combine these results to test the various
perturbative estimates for
next-to-leading order correction to Debye screening as
presented in the previous sections, in which
we had put in a magnetic screening mass by hand.

There exist also some results for SU(3), both without\cite{Gao,%
KLMPR} and with quarks\cite{KLPR},
but the corresponding lattice simulations have achieved
less precision and involve
smaller values of $T/T_c$. Because of the latter, one has to
expect more pronounced nonperturbative effects so that there is
less chance for
making contact with our perturbative considerations. Also, most of
the lattice results for the magnetic mass, which is a crucial input
in our perturbative estimates, have been obtained in pure SU(2).

We shall therefore concentrate on the results given in
Ref.~\cite{ILMPR}
and compare them with our various perturbative estimates.
The lattice simulations of magnetic screening in pure SU(2) gauge
theory\cite{mm} have given $m_{\mathrm m}=0.27(3) g^2T$,
which is consistent
with a recent semiclassical result by Bir\'o and M\"uller \cite{BiroM}
and also with a recent result from one-loop resummed gap equations
in a non-linear sigma model for the magnetic mass term\cite{OP}.
In what follows we shall plug in this value for the magnetic mass
into our perturbative result for the next-to-leading order
Debye mass (\ref{d}) with the above error and
evaluate it both strictly perturbatively and fully self-consistently.
As discussed at the end of
Sect.~2, in the strictly perturbative
case we only resum the classical (HTL)
Debye mass $m_0$ whereas in the second case we resum the full $m$,
solving the resulting
nonlinear equations for $m$ numerically. The variation
of the results will serve us as a crude measure for the uncertainties
of the resummed one-loop calculation.

In Ref.~\cite{ILMPR}, the Polyakov loop correlation has been measured
for lattice sizes $4\times 24^3$ at $g_0^2=1.43$ and $4\times 32^4$
at $g_0^2=1.33$, at distances of up to 8 lattice units.
In order to make contact with perturbation theory, the coupling
constant has to be renormalized. This is done at short distances,
that is, at the smallest
available distance of one lattice spacing, yielding
$g^2\approx 1.28$ and $g^2\approx 1.19$, respectively.
The smaller
value of the coupling,
which pertains to the larger lattice,
corresponds to the
rather high temperature $T/T_c\approx 7.8$ according to recent work
by Fingberg et al.\cite{FHK}.
With this renormalised value of $g$,
Eq.~(\ref{d}) yields for $m_{\mathrm m}/g^2T=0.27\pm0.03$
\begin{equation}
\frac{m}{m_0}=
\cases{1.22&$\mp0.02$\cr 1.36&$\mp0.04$\cr} ,
\label{mprp}\end{equation}
where the upper and lower lines correspond to perturbative and
self-consistent evaluation, respectively.
(With the larger value of $g$ the above values are almost unchanged.)

A screening mass has been extracted from the lattice data by
fitting the
unsubtracted
PLC to the mean-square behaviour of Eq.~(\ref{ms0}) at 
distances
of 4 to 8 lattice spacings. For both lattices, this
corresponds to values of $x\approx 0.9\ldots 1.8$. Although
these are not very large values, the mean-square approximation
is probably quite good in SU(2), since there the next important
correction term associated with 3 gluon exchange vanishes\cite{N1}.
The results reported in Ref.~\cite{ILMPR} are
\begin{equation}
\frac{m}{m_0}=\cases{1.11(5)&for $g^2\approx 1.28$\cr
  1.14(4)&for $g^2\approx 1.19$\cr}
\label{mlatt}\end{equation}
i.e., a significant
enhancement of the Debye screening mass over its classical value, which
is even more pronounced for the smaller coupling corresponding
to the higher temperature.\footnote{Debye screening
in SU(2) gauge theory
was also studied in Ref.~\cite{EKS},
but at somewhat smaller distances and not directly in terms of the
connected
Polyakov loop correlation.}

However,
for the moderate values of $x\approx 1\ldots2$
covered by the lattice calculations
these numbers cannot yet be compared with the perturbative results
which refer to the asymptotic behaviour. As we have seen in Sect.~5,
it is important to take the pre-exponential
screening functions into account when attempting to identify the
asymptotic screening mass. From Fig.~4 it is
clear that fitting the classical
function $(e^{-mr}/r)^2$ on a restricted range
of $x$ will underestimate\footnote{A different
approach to determining the
Debye screening mass, which is based on the transfer matrix formalism,
has recently been proposed in Ref.~\cite{EMN} and indeed leads
to somewhat
higher values than the conventional approach.} $m$.
In Fig.~6 the lattice data are compared with the perturbative
results of Sect.~5 for the screening function $F(x)\exp(-2xm/m_0)$.
Fitting straight lines in this logarithmic plot
gives an effective screening mass which reads
\begin{equation}
\frac{m_{eff.}}{m_0}=\cases{1.16(9)&PLC with $m_{\mathrm m}$ and
Debye mass from Eq.~(\ref{mprp})\cr
0.97&PLC without magnetic mass and $m=m_0$\cr}
\label{meffs}
\end{equation}
where the error gives the full variation caused by both
the quoted error of the magnetic mass and the spread from
a perturbative vs. self-consistent evaluation of the
one-loop formulae.
The PLC result without a magnetic mass thus leads to a
slightly diminished effective Debye mass as compared with the
classical value. On the other hand,
the increase of the Debye mass according to Eq.~(\ref{d})
is not completely compensated by the curvature of
$F(x)$
and leads to an effective Debye mass significantly larger than the
classical
value, in accordance with the lattice data. As concerns the
magnitude of the screening function, the lattice data lie in
between the perturbative results with and without a magnetic mass.

\begin{figure}                                                   
\centerline{ \epsfxsize=4.8in \epsfbox{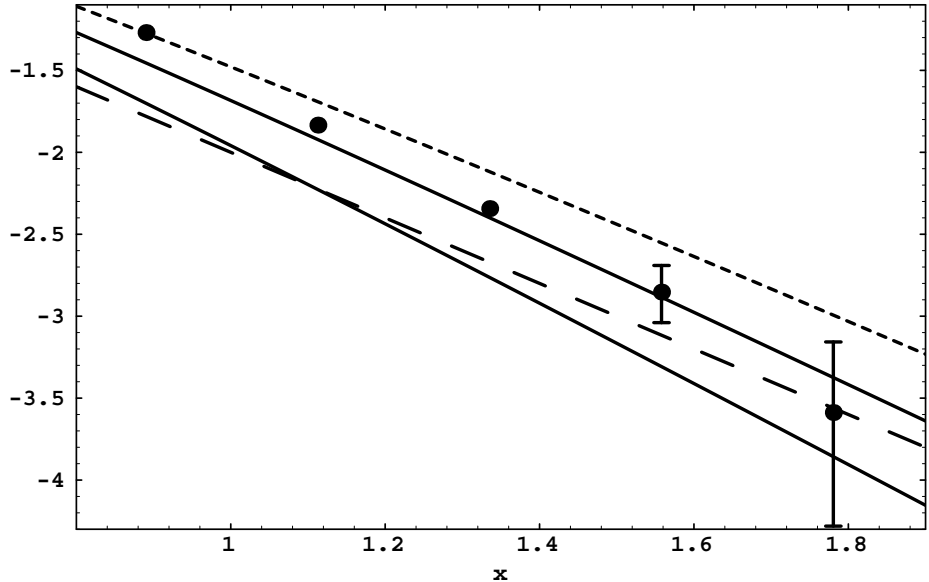} }
\caption[f6]{\label{f6}{The logarithm of the
screening function $F(x)\exp(-2xm/m_0)$ of the Polyakov loop
correlation
in the range of $x\equiv m_0 r$
covered by the lattice calculations of Ref.~\cite{ILMPR}.
The long-dashed straight line gives the classical result.
The short-dashed
curve gives the one-loop resummed result for the
mean-square contribution
prior to the introduction of a magnetic mass. The full lines
give the range of the results which include a magnetic mass
$m_{\mathrm m}=0.27 g^2T$.
The upper and lower full line
correspond to the upper and lower value of the Debye mass $m$ in
Eq.~(\ref{mprp}), respectively.
The lattice data for the connected Polyakov loop
correlation are given by the thick dots.
The error bars as reported in Ref.\cite{ILMPR}
are negligible except for the last two data points.
 }}       
\end{figure}                                                     

All of the above numbers depend crucially on the value of the
renormalized coupling constant associated with the bare one on the
lattice. So far we have followed the procedure of Ref.~\cite{ILMPR},
where the former was determined by matching the short-distance behaviour
to the tree-level result of the bare perturbation theory. Using instead
the resummed tree-level result which includes the leading-order
Debye mass necessarily
gives a different renormalized coupling because one cannot
go to smaller distances than one lattice spacing. In the case of
the larger lattice and higher temperature, the latter procedure
increases $g_0^2=1.33$ to $g^2_R\approx 1.53$, while previously the
renormalized coupling was diminished.
With this larger value of $g_R$, the results in Eq.~(\ref{meffs}) are
only insignificantly increased, yielding now
\begin{equation}
\frac{m_{eff.}}{m_0}=\cases{1.18(11)&PLC with $m_{\mathrm m}$ and
Debye mass according to Eq.~(\ref{d})\cr
0.98&PLC without magnetic mass and $m=m_0$\cr}
\label{meffsrr}
\end{equation}
whereas the lattice result (\ref{mlatt})
is reduced to $m/m_0=1.00(4)$. Apparently this is in better agreement
with the perturbative result prior to the introduction of
a magnetic mass, but by directly
comparing the complete screening functions with the
lattice data, redone in Fig.~7,
the case with a magnetic mass
and an enhanced Debye mass from Eq.(\ref{d}),
seems to be clearly favoured.

\begin{figure}                                                   
\centerline{ \epsfxsize=4.8in \epsfbox{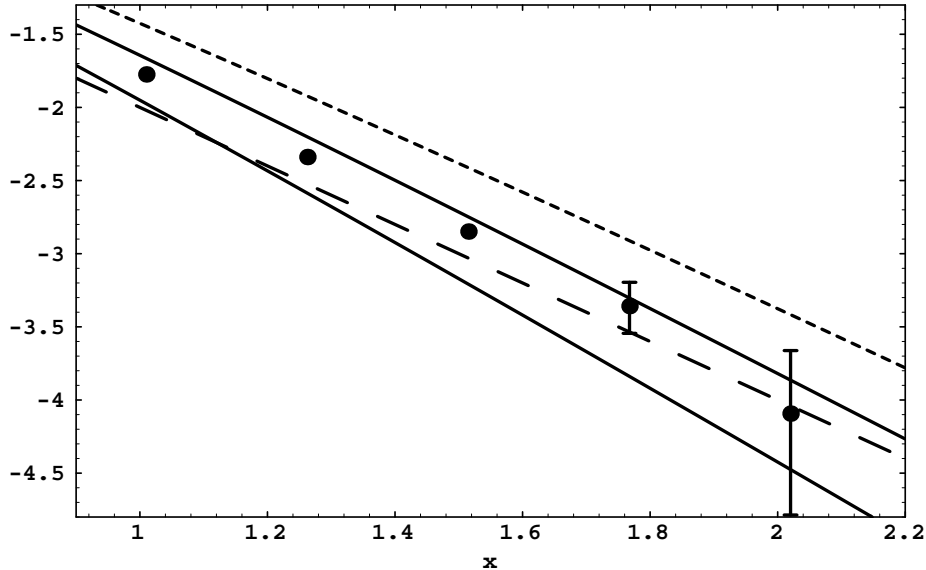} }
\caption[f7]{\label{f7}{Same as Fig.~6, but with a renormalized
coupling of $g^2\approx 1.53$, which is obtained by matching the
short-distance behaviour to the resummed (rather than bare) tree-level
result.}}       
\end{figure}                                                     

Finally we briefly turn also to
the case of an $A_0$ condensate, where we have encountered a rather
different analytic structure of the screening function. There
the theoretical uncertainty is much greater, because the
loop expansion parameter $1/c$ is always $\gtrsim1$ (with the
above parameters $1/c \approx 4$). In Sect.~6 we have
found an oscillatory behaviour that started out repulsive, but we
have argued that this result is not perturbatively stable. After
discarding this part of the result, there remained a contribution
from a
branch cut at $k=\pm 2im_{\mathrm m}$ instead of $\pm im$.
For very small coupling, this should be a distinctive feature, at least
at sufficiently large distances.
However, the lattice results involve only moderate values of $x$ and
furthermore, for the coupling constant that was used in the above
lattice simulations and the assumed magnetic mass,
we have $2m_{\mathrm m}/m=0.7(2)$. For large $x$ this suggests
a diminished Debye screening mass, but at the not so large $x$
for which lattice data exist, the effective screening mass
may be different. Indeed, the
screening function displayed in Fig.~5 has positive curvature
(in contrast to those of Fig.~4), so
for smaller $x$ the effective screening mass now becomes larger.
With the lattice parameters, the one-loop result yields
$m_{eff.}/m_0\approx1.05
$,
i.e. even a slight enhancement compared to the classical Debye mass,
thus going in the right direction.
However, as has been shown recently\cite{BoPr},
in the case of a nonvanishing
$A_0$ condensate the leading term in the PLC
is not the mean-square contribution of Eq.~(\ref{ms0}) but one
which involves only one power of the usual Yukawa-type
potential. Fitting the latter behaviour
to the lattice results of Ref.~\cite{ILMPR}
gives $m/m_0\approx 3.4(1)$ ($\approx 3.0(1)$ if the larger value of
the renormalized coupling is used), 
which is far off from the perturbative result.\footnote{A
single power of the Yukawa potential is
in much better shape just above the deconfinement transition\cite{EKS},
and it has been conjectured in Ref.~\cite{BoPr} that there might
be another, gauge-symmetry restoring phase
transition not far above the deconfinement temperature.}

In view of the largeness of the coupling and the uncertainty
associated with the treatment of the magnetic mass, one certainly
cannot expect
next-to-leading order results to be really accurate, but
the agreement of the lattice data with the screening function of
the symmetric scenario with a magnetic
mass $m_{\mathrm m}\sim g^2T/4$ and the enhanced
Debye mass according to Eq.~(\ref{d}) is
in fact surprisingly good.
The range of $x$ covered by the existing
lattice calculations is however
too small to really
pin down the actual asymptotic behaviour. Even the qualitatively
different asymptotics suggested by the
TAG result, which shows oscillatory behaviour with a repulsive power-law
tail\cite{BK}, would
manifest itself only for $x$ considerably greater than 2.
However, lattice simulations
at larger values of $x$ are probably prohibitively difficult, since
the statistical errors increase very quickly with $x$.
On the other hand,
progress with (resummed) perturbation theory seems only
possible through a better understanding of the nonperturbative
magnetic sector of thermal nonabelian gauge theories.
\acknowledgments

I would like to thank R. Baier, E. Braaten, W. Buchm\"uller,
J. Engels, J. Kapusta, P. Landshoff, D. Miller,
I. Montvay, S. Peign\'e, R. Pisarski, K. Redlich, T. Reisz, and S. Wong
for valuable discussions.
This research is supported in part by the EEC Programme ``Human Capital
and Mobility'', Network ``Physics at High Energy Colliders'', contract
CHRX-CT93-0357 (DG 12 COMA).

\appendix
\section*{}

In this appendix some details to the evaluation of the
PLC screening function are given when a magnetic screening mass
is included, supplementing the
calculations contained in the appendix of Ref.~\cite{N1}.

$F^{(b)}$, introduced in Sect.~5 and corresponding to the
function $f_I$ of Ref.~\cite{N1}, is given by
\begin{equation}
F^{(b)}=8\pi r e^{mr} \int \frac{d_3{\bf k}\,
e^{-i{\bf k}\cdot{\bf r}}}{({\bf k}^2+m^2)^2}
\left( \delta m^2-\delta\Pi_{00}(0,k) \right)
\end{equation}
where $d_3{\bf k} \equiv {d^{3-2\varepsilon}k}/
{(2\pi)^{3-2\varepsilon}}$
and
\begin{eqnarray}
&&\delta\Pi_{00}(k_0=0,{\bf k})
=g^2NT\int
d_3{\bf p}\,
\biggl\{\frac1{{\bf p}^2+m^2}
+\frac1{{\bf p}^2+m_{\mathrm m}^2} \nonumber\\
&&\qquad+\frac{2(m^2-m_{\mathrm m}^2/2
-{\bf k}^2)}{{\bf p}^2({\bf q}^2+m^2)}
+(\alpha-1) ({\bf k}^2+m^2)
\frac{{\bf p}^2+2{\bf p}{\bf k}}{({\bf p}^2
+m_{\mathrm m}^2)^2({\bf q}^2+m^2)}
 \biggr\},
\label{pi00mm}\end{eqnarray}
with
${\bf q}={\bf p}+{\bf k}$.

This leads to
\begin{eqnarray}
F^{(b)}&=& mr \frac{\delta m^2}{m^2}+\frac{g^2NT}{2\pi m}
\biggl\{
\frac{(m+m_{\mathrm m})r}2 + 4\pi m
(1-\alpha) \int d_3{\bf l}\,
\frac{(1-e^{i{\bf l}\cdot{\bf r}})}{({\bf l}^2+m_{\mathrm m}^2)^2}
\\ \nonumber
&&+\frac{i}{\pi} \int_C dz\, e^{-mrz}
\frac{z^2+2z+2-\mu^2/2}{z^2(2+z)^2}
\ln\frac{\mu-z}{2+\mu+z} \biggr\},
\label{fi}\end{eqnarray}
where the contour $C$ encircles the positive real axis
and $\mu\equiv m_{\mathrm m}/m$.

The contour integral in (\ref{fi})
receives contributions from a pole at $z=0$ and
a branch cut starting at $z=\mu$, which thus is separated from the pole
by the finite magnetic mass.
The contribution from the pole reads
\begin{equation}
 i\pi \biggl\{ (1-\mu^2/4) \left[
-mr \,\ln\frac{\mu}{2+\mu} - \frac1{\mu}-\frac1{2+\mu} \right]
 +\frac{\mu^2}4\ln\frac{\mu}{2+\mu} \biggr\},
\end{equation}
and the one from the cut
\begin{eqnarray}
&& i\pi\biggl\{(1-\mu^2/4) \left[ e^{-m_{\mathrm m} r}
\left(\frac1{\mu}+\frac1{2+\mu}\right)+mr\,
{\rm Ei}(-m_{\mathrm m}r)+mr\,e^{2mr}\,{\rm Ei}(-(2+\mu)
mr) \right] \\ \nonumber
&& \qquad -\frac{\mu^2}4 {\rm Ei}(-m_{\mathrm m}r)+\frac{\mu^2}4
e^{2mr} \, {\rm Ei}(-(2+\mu) mr)\biggr\}.
\end{eqnarray}

In general covariant gauge and with a magnetic mass,
$F^{(c)}$, which corresponds to the function $f_{II}$ of Ref.~\cite{N1},
reads
\begin{equation}
F^{(b)}=g^2NT
\left\{ -2(1-\alpha)I_2
+(4\pi r)^2e^{2mr} \left[ 4 I_3 - I_4 -2 I_5 \right]\right\}
\label{fii}\end{equation}
with
\begin{eqnarray}
I_3 &=& \int \frac{d_3{\bf k}_1\,d_3{\bf k}_2\,d_3{\bf l}\,
e^{-i({\bf k}_1-{\bf k}_2)\cdot{\bf r}}}{
({\bf k}_1^2+m^2)[({\bf k}_1+{\bf l})^2+m^2]
({\bf k}_2^2+m^2)({\bf l}^2+m_{\mathrm m}^2)} \\
I_4 &=& \int d_3{\bf k}\, e^{-i{\bf k}\cdot{\bf r}} \left[
\int\frac{d_3{\bf l}}{[({\bf l}+{\bf k})^2+m^2]({\bf l}^2+m^2)}
\right]^2 \\
I_5 &=& \int \frac{d_3{\bf k}\,e^{-i{\bf k}\cdot{\bf r}}
({\bf k}^2+2m^2-m_{\mathrm m}^2/2) \,
d_3{\bf p}\,d_3{\bf q}}{[({\bf p}-{\bf q})^2
+m_{\mathrm m}^2]({\bf p}^2+m^2)({\bf q}^2+m^2)
[({\bf p}+{\bf k})^2+m^2][({\bf q}+{\bf k})^2+m^2]}.
\end{eqnarray}

The gauge-parameter dependent parts of
$F^{(c)}$ exactly cancel those of $F^{(b)}$. The integral appearing
therein yields
\begin{equation}
I_2\equiv\int d_3{\bf l} \frac{(1-
e^{i{\bf l}\cdot{\bf r}})}{({\bf l}^2+m_{\mathrm m}^2)^2}
=\frac1{8\pi m_{\mathrm m}}\left( 1-e^{-m_{\mathrm m}r} \right),
\end{equation}
so even separately
these terms do not contribute to the linear terms in $r$ that constitute
the next-to-leading order shift of the Debye screening mass if the
unphysical gauge degrees of freedom are made massive; if they are
kept massless, $I_2\to r/(8\pi)$,
and these linear contributions cancel upon adding $F^{(b)}$
and $F^{(c)}$.

In Ref.~\cite{N1} all of the above integrals were evaluated exactly
in the case of a vanishing magnetic mass. With a finite magnetic mass,
$I_3$ can still be calculated completely, which yields
\begin{eqnarray}
I_3&=& \frac{i e^{-2mr}}{(4\pi r)^2 8\pi^2 m}
\int_C \frac{dz\,e^{-mrz}}{z(2+z)}\ln\frac{\mu-z}{2+\mu+z}\\ \nonumber
&=&\frac{e^{-2mr}}{(4\pi r)^2 8\pi m}
\left[ {\rm Ei}(-m_{\mathrm m}r)-e^{2mr} \, {\rm Ei}(-(2+\mu) mr)-
\ln\frac{\mu}{2+\mu} \right].
\end{eqnarray}

$I_4$ is unchanged by the introduction of the magnetic mass
and reads
\begin{equation}
I_4=\frac{e^{-2mr}}{(4\pi r)^2} \frac{r}{4\pi} g(2mr),
\end{equation}
with $g$ as defined in Eq.~(\ref{gx}).

$I_5$, with a finite magnetic mass, can be brought into the form
\begin{equation}
I_5 = \int d_3{\bf k}\, e^{-i{\bf k}\cdot{\bf r}}
({\bf k}^2+2m^2-m_{\mathrm m}^2/2) J(k)
\end{equation}
where $J(k)$ is given by the double integral
\begin{eqnarray}
J(k)&=& \frac1{32\pi^2} \int_0^1
\frac{dx}{\sqrt{k^2 x(1-x)+m^2}} \\ \nonumber
&\times& \int_0^1 \frac{ dy (1-2xy)}{[k^2+4(m^2+yW)]
\sqrt{k^2 xy(1-xy)+m^2+yW}} \\ \nonumber
\mbox{where}&&W(k,x)\equiv m_{\mathrm m}^2
+2m_{\mathrm m}\sqrt{k^2 x(1-x)+m^2}.
\end{eqnarray}

With $m_{\mathrm m}=W=0$,
\begin{equation}
J(k)=\frac1{16\pi^2 k^2(k^2+4m^2)}\ln\frac{k^2+4m^2}{4m^2},
\end{equation}
which leads to
\begin{eqnarray}
I_5 &=& \frac{e^{-2mr}}{(4\pi r)^2}\frac{r}{8i\pi^2}
\int_C \frac{dz\,e^{-2mrz}(1+4z+2z^2)}{z(1+z)(2+z)}\ln[-z(2+z)]
\\ \nonumber
&=& -\frac{e^{-2mr}}{(4\pi r)^2}\frac{r}{8\pi}
\left[ \ln(mr)+\gamma+2e^{2mr}\,{\rm Ei}(-2mr)+
e^{4mr}\,{\rm Ei}(-4mr) \right].
\end{eqnarray}

As has been pointed out first in Ref.~\cite{BN}, a finite magnetic mass
removes the pole at $k^2=-4m^2$ (corresponding to $z=0$) and replaces it
by neighbouring branch singularities at $k^2=-4m^2$ and
$k^2=-(2m+m_{\mathrm m})^2$.
In the limit $\kappa^2=-k^2\to 4m^2$, $J(k)$ yields
\begin{equation}
J(k)\Big|_{k^2\to-4m^2} \sim
\frac1{(16\pi m m_{\mathrm m})^2}\ln\frac{2m-\kappa}{4m}
\ln\frac{4m^2-\kappa^2}{4m_{\mathrm m}^2},
\end{equation}
which leads to the large distance behaviour
\begin{equation}
I_5 \sim -\frac{e^{-2mr}}{(4\pi r)^2}\frac{m}{4\pi m_{\mathrm m}^2}
(\ln2m_{\mathrm m}r+\gamma),
\end{equation}
up to terms that are down by powers of $m_{\mathrm m}/m$.

Putting everything together leads to Eq.~(\ref{fmx}), where the
full function $I_5$ has been approximated in terms of the function
$\ell(x)$ introduced in Eq.~(\ref{ell}).


\begin{references}
\bibitem{FT}
J. Frenkel and J. C. Taylor, {\it Nucl. Phys.} {\bf B334} (1990) 199.

\bibitem{BP}
E. Braaten and R. D. Pisarski,
{\it Phys. Rev. Lett.} {\bf64} (1990) 1338;
{\it Nucl. Phys.} {\bf B337} (1990) 569.

\bibitem{Silin}
V. P. Silin, {\it Zh. Eksp. Teor. Fiz. } {\bf 38} (1960) 1577
[{\it Sov. Phys.} {\bf 11} (1960) 1136];
U. Heinz, {\it Ann. Phys. (N.Y.)} {\bf 161} (1985) 48; {\it ibid.}
{\bf 168} (1986) 148;
J.-P. Blaizot and E. Iancu, {\it Nucl. Phys.} {\bf B417} (1994) 608;
P. F. Kelly, Q. Liu, C. Lucchesi and C. Manuel, {\it Phys. Rev. Lett.}
{\bf 72} (1994) 3461.

\bibitem{forwscatt}
G. Barton, {\it Ann. Phys. (N.Y.)} {\bf200} (1990) 271;
J. Frenkel and J. C. Taylor, {\it Nucl. Phys.} {\bf B374} (1992) 156.

\bibitem{Damping}
E. Braaten and R. D. Pisarski, {\it Phys. Rev.} {\bf D42} (1990) 2156;
{\it ibid.} {\bf D46} (1992) 1829;
R. Kobes, G. Kunstatter and K. Mak,
{\it Phys. Rev.} {\bf D45} (1992) 4632.

\bibitem{HS}
H. Schulz, {\it Nucl. Phys.} {\bf B413} (1994) 353.

\bibitem{Parwani}
R. R. Parwani, {\it Phys. Rev.} {\bf D45} (1992) 4695.

\bibitem{SED}
U. Kraemmer, A. K. Rebhan and H. Schulz, DESY-report 94-034 (1994), to
appear in Ann. Phys. (N. Y.).

\bibitem{Damping2}
R. D. Pisarski, {\it Phys. Rev. Lett.} {\bf63} (1989) 1129;
V. V. Lebedev and A. V. Smilga, {\it Physica} {\bf A181} (1992) 187;
C. P. Burgess and A. L. Marini, {\it Phys. Rev.} {\bf D45} (1992) R17;
A. Rebhan, {\it ibid.} {\bf D46} (1992) 482;
T. Altherr, E. Petitgirard and T. del Rio Gaztelurrutia,
{\it Phys. Rev.} {\bf D47} (1993) 703.

\bibitem{Damping3}
R. Baier, H. Nakkagawa and A. Ni\'egawa, {\it Can. J. Phys.} {\bf 71}
(1993) 205; R. Pisarski, {\it Phys. Rev.} {\bf D47} (1993) 5589;
S. Peign\'e, E. Pilon and D. Schiff, {\it Z. Phys.} {\bf C60} (1993) 455.

\bibitem{AKR}
A. K. Rebhan, {\it Phys. Rev.} {\bf D48} (1993) R3967.

\bibitem{Satz}
T. Matsui and H. Satz, {\it Phys. Lett.} {\bf 178B} (1986) 416.

\bibitem{CM}
F. Cornu and Ph. A. Martin, {\it Phys. Rev.} {\bf A44} (1991) 4893.

\bibitem{BK}
R. Baier and O. K. Kalashnikov, {\it Phys. Lett.} {\bf B328} (1994) 450.

\bibitem{PW}
S. Peign\'e and S. M. H. Wong, preprint LPTHE-Orsay 94/46 (1994).

\bibitem{BN}
E. Braaten and A. Nieto, preprint NUHEP-TH-94-18 (1994).

\bibitem{Weldon}
H. A. Weldon, {\it Phys. Rev.} {\bf D26} (1982) 1394.

\bibitem{Kapusta}
J. I. Kapusta, {\it Finite Temperature Field Theory}
(Cambridge University Press, Cambridge, 1989).

\bibitem{T}
T. Toimela, {\it Z. Phys.} {\bf C27} (1985) 289.

\bibitem{KKR}
R. Kobes, G. Kunstatter and A. Rebhan,
{\it Phys. Rev. Lett.} {\bf64} (1990) 2992;
{\it Nucl. Phys.} {\bf B355} (1991) 1.

\bibitem{Fradkin}
E. S. Fradkin, 
{\it Proc. of the Lebedev Institute} {\bf 29} (1965) 6.

\bibitem{AE}
P. Arnold and O. Espinosa, 
{\it Phys. Rev.} {\bf D47} (1993) 3546.

\bibitem{GMB}
M. Gell-Mann and K. A. Brueckner, {\it Phys. Rev.} {\bf106} (1957) 364.

\bibitem{BKS}
R. Baier, G. Kunstatter and D. Schiff, {\it Phys. Rev.} {\bf D45}
(1992) 4381; {\it Nucl. Phys.} {\bf B388} (1992) 287.

\bibitem{BKSC}
A. Rebhan, {\it Phys. Rev.} {\bf D46} (1992) 4779.

\bibitem{Sintra}
A. K. Rebhan, to appear in the proceedings of the NATO Advanced
Research Workshop ``Electroweak Physics and the Early Universe'',
March 1994, Sintra, Portugal.

\bibitem{Linde}
A. D. Linde, {\it Phys. Lett.} {\bf B96} (1980) 289.

\bibitem{GPY}
D. Gross, R. Pisarski and L. Yaffe,
{\it Rev. Mod. Phys.} {\bf53} (1981) 43.

\bibitem{DJ}
L. Dolan and R. Jackiw, {\it Phys. Rev.} {\bf D9} (1974) 3320.

\bibitem{KK}
K. Kajantie and J. Kapusta, {\it Ann. Phys. (N.Y.)} {\bf160} (1985) 477.

\bibitem{HKT}
U. Heinz, K. Kajantie and T. Toimela, {\it Ann. Phys. (N. Y.)}
{\bf 176} (1987) 218.

\bibitem{KKRS}
U. Kraemmer, M. Kreuzer, A. Rebhan and H. Schulz, {\it Lect. Notes in
Phys.} {\bf 361} (1990) 285.

\bibitem{CCM}
S. Carracciolo, G. Curci and P. Menotti, {\it Phys. Lett.}
{\bf 113B} (1982) 311.

\bibitem{KPVL}
K. James and P. V. Landshoff, {\it Phys. Lett.} {\bf B251} (1990) 167.

\bibitem{HN}
M. Kreuzer and H. Nachbagauer, {\it Phys. Lett.} {\bf B271} (1991) 155;
H. Nachbagauer, {\it Z. Phys.} {\bf C56} (1992) 407.

\bibitem{N1}
S. Nadkarni, {\it Phys. Rev.} {\bf D33} (1986) 3738.

\bibitem{McLS}
L. D. McLerran and B. Svetitsky, {\it Phys. Rev.} {\bf D24} (1981) 450.

\bibitem{GR}
I. S. Gradshteyn and I. M. Ryzhik, {\it Table of Integrals, Series,
and Products} (Academic Press, Orlando 1980).

\bibitem{A0}
R. Anishetty, {\it J. Phys. G} {\bf 10} (1984) 423;
J. Polonyi and H. W. Wyld, Illinois Report No. ILL-TH-85-23 (1985);
K. J. Dahlem, {\it Z. Phys.} {\bf C29} (1985) 553;
J. Boh\'a\v cik, {\it Phys. Rev.} {\bf D42} (1990) 3554;
M. Oleszczuk, in {\it Hot Summer Daze}, eds. A. Gocksch and R. Pisarski
(World Sci., Singapore 1992).

\bibitem{N2}
S. Nadkarni, {\it Phys. Rev.} {\bf D34} (1986) 3904.

\bibitem{ILMPR}
A. Irb\"ack, P. LaCock, D. Miller, B. Petersson and T. Reisz,
{\it Nucl. Phys.} {\bf B363} (1991) 34.

\bibitem{Gao}
M. Gao, Phys. Rev. D {\bf41} (1990) 626.

\bibitem{KLMPR}
L. K\"arkk\"ainen, P. LaCock, D. E. Miller, B. Petersson and T. Reisz,
{\it Phys. Lett.} {\bf B282} (1992) 121.

\bibitem{KLPR}
L. K\"arkk\"ainen,
P. LaCock, B. Petersson and T. Reisz, Nucl. Phys. {\bf B395} (1993)
733.

\bibitem{mm}
A. Billoire, G. Lazarides and Q. Shafi, {\it Phys. Lett.} {\bf 103B}
(1981) 450; T. A. DeGrand and D. Toussaint, {\it Phys. Rev.} {\bf D25}
(1982) 526.

\bibitem{BiroM}
T. S. Bir\'o and B. M\"uller, {\it Nucl. Phys.} {\bf A561} (1993) 477.

\bibitem{OP}
O. Philipsen, to appear in the proceedings of
the NATO Advanced
Research Workshop ``Electroweak Physics and the Early Universe'',
March 1994, Sintra, Portugal.

\bibitem{FHK}
J. Fingberg, U. Heller and F. Karsch,
Nucl. Phys. {\bf B392} (1993) 493.

\bibitem{EKS}
J. Engels, F. Karsch and H. Satz,
{\it Nucl. Phys.} {\bf B315} (1989) 419.

\bibitem{EMN}
J. Engels, K. Mitrjushkin and T. Neuhaus, 
{\it Nucl. Phys. B (Proc. Suppl.)} {\bf 34} (1994) 298.

\bibitem{BoPr}
J. Boh\'a\v cik and P. Pre\v snajder, {\it Phys. Lett.} {\bf B332}
(1994) 366.
\end{references}
\end{document}